\documentclass[english,aps,preprint]{revtex4}
\usepackage[T1]{fontenc}
\usepackage[latin9]{inputenc}
\usepackage{mathrsfs}
\usepackage{amsmath}
\usepackage{graphicx}

\makeatletter
\@ifundefined{textcolor}{}
{%
 \definecolor{BLACK}{gray}{0}
 \definecolor{WHITE}{gray}{1}
 \definecolor{RED}{rgb}{1,0,0}
 \definecolor{GREEN}{rgb}{0,1,0}
 \definecolor{BLUE}{rgb}{0,0,1}
 \definecolor{CYAN}{cmyk}{1,0,0,0}
 \definecolor{MAGENTA}{cmyk}{0,1,0,0}
 \definecolor{YELLOW}{cmyk}{0,0,1,0}
}

\makeatother

\usepackage{babel}
\begin{document}

\title{Entropic Entanglement: Information Prison Break }

\author{Alexander Y. Yosifov}
\email{alexanderyyosifov@gmail.com}

\author{Lachezar G. Filipov}
\email{lfilipov@mail.space.bas.bg}

\affiliation{Space Research and Technology Institute, Bulgarian Academy of Sciences}
\begin{abstract}
We argue certain nonviolent local quantum field theory (LQFT) modification
considered at the global horizon ($r=2M$) of a static spherically-symmetric
black hole can lead to adiabatic leakage of quantum information in
the form of Hawking particles. The source of the modification is (i)
smooth at $r=2M$ and (ii) rapidly vanishing at $r\gg2M$. Furthermore,
we restore the unitary evolution by introducing extra quanta which
departs slightly from the generic Hawking emission without changing
the experience of an infalling observer (no drama). Also, we suggest
that a possible interpretation of the Bekenstein-Hawking bound as
entanglement entropy may yield a nonsingular dynamical horizon behavior
described by black hole thermodynamics. Hence by treating gravity
as a field theory, and considering its coupling to the matter fields
in the Minkowski vacuum, we derive the conjectured fluctuations of
the background geometry of a black hole.
\end{abstract}
\maketitle

\section*{1. Introduction}

It has been argued in {[}1{]} that black holes are not black at all.
Rather, in a semi-classical approximation they are shown to be hot
bodies that emit thermal radiation with an inverse temperature of
$2\pi$. The fate of information fallen into a black hole is still
under debate. Hawking's original proposal of loss of information,
and thus pure-to-mixed state evolution has been strongly opposed {[}2-7{]}
as it implies violation of quantum-mechanical unitarity. Resolution
of the information paradox within the current nomenclature does not
seem to be a fruitful endeavor. Instead, we have focused on modifying
already existing principles. Abandoning locality above the Planck
mass $(m_{p})$ for instance, appears as a promising and somewhat
more conservative approach. Significant theoretical support for fundamental
non-locality has come from AdS/CFT duality and cosmology {[}8-12{]}.
One of the authors {[}12{]} has shown that in the extreme conditions
of the early universe (superplanckian energies) non-locality plays
an essential role for explaining the origin of the cosmological principle.

Following the theoretical evidence, in the current paper we embrace
the notion of locality as an effective field theory, manifesting in
weak gravitational dynamics. Based on that assumption we propose a
framework, featuring a modification of local quantum field theory
as defined on the global horizon ($r=2M$) in order to provide a nonviolent
mechanism for taking the Hawking quanta, hence quantum information
out of the hole, and restore unitarity. The suggested LQFT modification
leads to weak (nonviolent) quantum effects which manifest in brief
non-local phenomena. Although weak, the perturbative nature of the
effects makes them significant in the course of a black hole's lifetime
$\mathcal{O}(M^{3}).$ The current model also predicts minor deviations
from the generic Hawking emission sufficient to restore the unitary
evolution without causing drama for an observer in free fall.

By treating gravity in a black hole background metric (Minkowski space)
as a field theory $(\mathit{graviton})$ we derive the microscopic
origin of the conjectured Planckian-amplitude horizon oscillations
{[}13{]}. In a previous paper we approached the phenomena classically
by deriving the oscillations from perturbation theory (see Ref. {[}13{]}).
The results we obtain in the particular work may be considered as
a microscopic origin of the stretched horizon in observer complementarity
{[}18{]}. Furthermore, the effects of the source of the LQFT modification
are shown to respect the equivalence principle, and also to rapidly
vanish at large $r$.

The paper is organized as follows. In Section 2 we put forward the
LQFT revisions and show how the coupling between the graviton and
the matter fields in Minkowski space can lead to information escape.
In Section 3 we provide a quantum theory derivation of the proposed
horizon oscillations by treating gravity in the vicinity of a black
hole as field theory.

\section*{2. Non-local information release}

The particular modification of LQFT, which we propose, comes from
localized and brief violations of locality, yielded by \textquotedbl{}strong\textquotedbl{}
fluctuations of the $graviton$ that come from its coupling to the
matter fields in Minkowski space. We also put forward a gedanken experiment
which involves a pair of strings put on both sides of the future Rindler
horizon to illustrate that the equivalence principle holds of an observer
in free fall. 

Based on the thermal spectrum of the emitted radiation and black hole
thermodynamics (Second Law, in particular) we now take for granted
the proportionality between black hole entropy and horizon area. Namely,
the entropy of a black hole is one fourth of the area of the event
horizon in Planck units 

\begin{equation}
S_{BH}=\frac{A}{4}
\end{equation}
The geometric entropy bound is deeply rooted in holography, and further
generalized in the Bousso bound {[}14{]}. Yet, its origin has not
been fully explained. 

\subsection*{2.1 Geometric entropy = entanglement entropy}

We begin by showing how the Bekenstein formula can be derived from
entanglement entropy, and later how this can yield the small departures
from LQFT needed to carry the quantum information out of the black
hole.

Bianchi has shown {[}15{]} that by considering the correlations between
gravity and matter fields in the near-horizon region one can reproduce
the Bekenstein-Hawking bound Eq. (1). The obtained equality has been
shown to be universal and independent of the number of field $species$.
Hence Bianchi's derivation of the equality $S_{ent}=S_{BH}$, where
$S_{ent}$ is entanglement entropy, shows that quantum entanglement
is the fundamental origin of the Bekenstein entropy bound. Let us
further clarify that.

Imagine we have a Schwarzschild black hole in a pure state $\left|\varPsi\right\rangle $
with metric 

\begin{equation}
ds^{2}=-(1-2M/r)dt^{2}+(1-2M/r)^{-1}dr^{2}+r^{2}(d\theta^{2}+sin^{2}\theta d\phi^{2})
\end{equation}
where the singularity is at $r=0$ and the global horizon is at $r=2M$.

Here a black hole event horizon provides a perfect entangling surface
as it naturally causally disconnects the interior and exterior regions

\begin{equation}
\mathcal{H}=\mathcal{H}_{A}\otimes\mathcal{H}_{B}
\end{equation}
where $\mathcal{H}_{A}$ is the interior region $(r<2M)$ and $\mathcal{H}_{B}$
is the exterior $(r>2M)$. The dimensionality of $\mathcal{H}_{A}$
is given as the logarithm of the internal degrees of freedom.

The pure state of the complete system is given by the product of the
two subsystems

\begin{equation}
\left|\varPsi\right\rangle =\sum_{i}\left|A_{i}\right\rangle \otimes\left|B_{i}\right\rangle 
\end{equation}

$\vphantom{}$

$\vphantom{}$

With a corresponding density matrix

\begin{equation}
\rho=\left|\varPsi\right\rangle \left\langle \varPsi\right|
\end{equation}

The pure state of the complete system may be decomposed as

\begin{equation}
\left|\varPsi\right\rangle \rightarrow\rho_{A}+\rho_{B}
\end{equation}
where $\rho_{A}$ and $\rho_{B}$ denote the reduced density matrices
of the corresponding subsystems $A$ and $B$, respectively. Note
that in a black hole background the initial state cannot be trivially
reproduced by the thermal density matrices; some of the information
concerning $\left|\varPsi\right\rangle $ is found in the entanglement
between the two subsystems across the entangling surface.

Consider the Minkowski vacuum in the region near the black hole which
is bounded by a local Rindler horizon $\mathcal{H}^{+}$. The complementary
$\mathit{left}$ and $\mathit{right}$ Rindler wedges, described by
the Hilbert spaces $\mathcal{H}_{A}$ and $\mathcal{H}_{B}$, respectively,
are given in terms of thermal density matrices, Eq. (6). The complete
vacuum state $\rho_{0}$ is due to the entangling between the field
theories defined on both sides of the horizon ($L$ and $R$ Rindler
wedges)

\begin{equation}
\rho_{0}=\sum_{i}e^{-\beta}\left|E_{i}^{A}\right\rangle \otimes\left|E_{i}^{B}\right\rangle 
\end{equation}
where $\left|E_{i}^{A}\right\rangle $ and $\left|E_{i}^{B}\right\rangle $
are the eigenstates associated with the wedges, and $\beta$ denotes
the inverse temperature.

In the vacuum state every mode on the $left$ Rindler wedge is entangled
with the corresponding mode on the $right$ wedge. The particular
entanglement normalizes the stress-energy tensor at $r=2M$, and hence
an infalling observer does not feel anything out of the ordinary.
The stress tensor normalization provides a smooth transition between
the two distinct causal patches. Thus the entanglement entropy is
proportional to the entangling surface (horizon). As a result, only
modes very close to the global horizon contribute to the entropy of
the system.

Bianchi's derivation of the equality $S_{ent}=S_{BH}$ strongly advocates
the entanglement origin of the geometric entropy. Considering the
correlations between gravity and matter fields in Minkowski space
suggests we can treat gravity as a field theory $(graviton)$. In
particular, we argue that by treating gravity in Minkowski space in
terms of quantum field theory in black hole background we can obtain
the desired modification of local quantum field theory in the vicinity
of the horizon, and thus present a framework for gradual release of
quantum information.

\subsection*{2.2 Gravity as a field theory}

Suppose we assign a time-dependent Killing frequency to the $graviton$
with respect to the background metric. The correlated quantum fields
(gravity and matter) in the near-horizon region oscillate rapidly,
and have radial dependence with respect to the Rindler horizon $\mathcal{H}^{+}$.
Generally, the quantum fluctuations of the matter fields near the
horizon get \textquotedbl{}amplified\textquotedbl{} by the black hole's
internal degrees of freedom and an inertial observer with a measuring
apparatus in that region measures $\left\langle N_{i}\right\rangle =\left\langle 0\right|N\left|0\right\rangle $,
where the expectation value of $N$ is non-zero (Hawking process).

We wish to focus on the black hole metric back-reaction from the $graviton$
fluctuations in two cases (i) Minkowski space, and (ii) the vicinity
of the horizon. Since we consider locality in effective field theory
to be a constraint imposed by the background geometry, we wish to
examine how fluctuations of the gravitational field above a threshold
$\lambda$ affect it.

We believe that strong quantum fluctuations of the gravitational field
as considered $\mathit{onto}$ the horizon ($r=2M$) will cause \textquotedbl{}disturbances\textquotedbl{}
in the background metric, and thus in the imposed by it locality.
Consider the following gedanken experiment. Imagine we have a static
black hole, Eq. (2), Figure 1.

\begin{figure}
\includegraphics[scale=0.65]{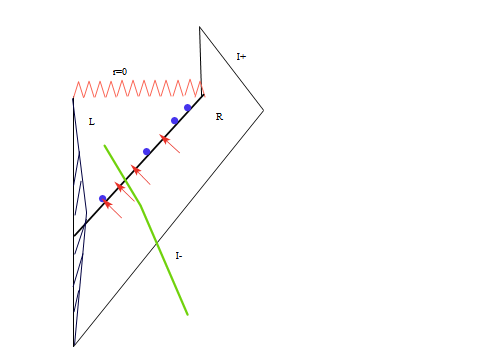}\caption{Penrose diagram depicting the matching between the randomly embedded
qubits (blue dots) and strong fluctuations (red arrows). The wave-like
red line is the singularity ($r=0$), the solid black line is the
global horizon ($r=2M$), and the green line is the worldline of an
infalling observer. $L$ and $R$ stand for $left$ and $right$ Rindler
wedge, respectively. $\mathcal{L}^{-}$ and $\mathcal{L}^{+}$ denote
past and future null infinity, respectively.}
 
\end{figure}
For convenience when describing the strong graviton fluctuations we
will consider a time slice on which the mass of the hole is time-translation
invariant. Hence the influx of matter exactly matches the emission
of Hawking particles to future null infinity. In the particular case
the emission should be thought of as a statistical phenomena which
solely depends on the internal Hilbert space $\mathcal{H}_{A}$. However,
small deviations may appear due to the random nature of the fluctuations
(excitations). Also, consider the following assumptions: (i) the future
Rindler horizon $\mathcal{H}^{+}$ is given in terms of null light
rays which neither get inside the black hole nor get emitted to asymptotic
infinity, $r<2M\:<\:\mathcal{H}^{+}\:<\:r>2M$, (ii) black holes act
as fast scramblers {[}26{]} and information is found in the emitted
quanta, and (iii) scrambled information $\mathit{need\:not}$ be embedded
uniformly across the horizon as this would make $log\,R\:<R\,log\,R$,
and as a result, an outside observer would be able to verify a violation
of the linearity of quantum mechanics. Note that in a recent paper
{[}16{]} we reproduced (with different initial assumptions and approach)
the results of Page {[}17{]} regarding information distribution onto
the horizon.

That being said, suppose we place two corresponding strings on both
sides of $\mathcal{H}^{+}$. Imagine we have one string {[}25{]} with
randomly placed qubits on it located very close to the global horizon
$r<2M$. Particularly, suppose the randomly distributed scrambled
qubits {[}26{]} in the left Rindler wedge, Fig.1, are interpreted
as switches turned $\mathit{on}$. Similarly, the string on the corresponding
right wedge ($r>2M$) is said to have statistical $graviton$ excitations
which can also be interpreted as switches, however, with one subtle
distinction. We treat the fluctuations as switches turned $\mathit{on}$
if their energy density exceeds a certain threshold $\lambda$. More
precisely:

(A) We think of a fluctuation of the $graviton$ as strong (switch
$\mathit{on}$) if, when considered at asymptotic spatial infinity,
its local energy density can polarize the vacuum, and thus produce
a particle $a_{i}^{\dagger}\left|0\right\rangle =\left|x\right\rangle $.
Therefore we can associate the strong fluctuations in a given spacetime
region with the expectation value for particle production $\left\langle N_{i}\right\rangle $
in that region. Generally

\begin{equation}
\sum_{i=1}^{N}\int_{\Sigma}\varphi_{strong}=\left\langle N_{i}\right\rangle .
\end{equation}
where $\varphi_{strong}>\lambda$.

(B) The fluctuations below that threshold, thus $\varphi_{weak}<\lambda$
are considered weak (switch $\mathit{off}$). Namely, $a_{i}\left|0\right\rangle =\left|0\right\rangle $. 

Ref. {[}28{]} provides a good description of the pair of corresponding
strings, acting on both sides of the Rindler horizon; namely for an
infalling observer we get

\begin{equation}
\int_{\Sigma}\mathcal{O}_{L}\mathcal{\phi}_{LR}\mathcal{O}_{R}
\end{equation}
where the operators $\mathcal{O}_{L}$ and $\mathcal{O}_{R}$ are
the corresponding ones for the $left$ and $right$ Rindler wedges,
respectively; and $\phi_{LR}$ is the source (graviton) which depends
on the internal degrees of freedom of the black hole.

We are interested in the quantum effects which arise when there is
a correspondence in the relative states of the occupation numbers
on both sides of $\mathcal{H}^{+}$.That is a strong fluctuation in
the $R$ region and a qubit in the $L$ region, where both are on
the same Cauchy surface
\begin{equation}
\left|00\right\rangle +\left|11\right\rangle 
\end{equation}
Let us consider a portion of the Rindler-like horizon $R_{H}$, where
$R_{H}\ll A_{H}$, and $A_{H}$ is horizon area

\begin{equation}
\int_{R_{H}}\delta\phi_{graviton}^{(x)}
\end{equation}
where the $graviton$ is $\phi$, and $\delta$ denotes the variations
of the field. 

So a strong fluctuation $(\varphi>\lambda)$ corresponding to an embedded
information (switch $\mathit{on}$), Fig. 1, would briefly disturb
the local dynamics of the background metric, and thus yield $[\phi(x),\,\phi(y)]=0$.
As a result, the corresponding mode, placed on the inner string ($left$
Rindler wedge), will be radiated to $\mathit{\mathscr{I}^{+}}$ as
a low $T$ Hawking particle. The \textquotedbl{}brief\textquotedbl{}
disturbance should be of order the lifetime of the fluctuation $\varphi$,
and should not lead to significant changes in the background metric.
Since the modifications are considered $\mathit{onto}$ $r=2M$, the
near-horizon physics is consistent with the postulates of complementarity
(Postulate II, in particular) {[}18{]}, and thus the non-local effects
are completely non-violent for an observer close to the horizon. For
instance, suppose that Alice is in the Minkowski vacuum carrying a
measuring apparatus. When she performs measurements, she will see
the typical effective field theory correlations between the exterior
and interior of the black hole. Complementary, Bob, being far away
from the black hole will not see any significant deviations from the
semi-classical Hawking framework.

We now examine the results from treating gravity as a field theory,
and considering the fluctuations which arise from its universal coupling
to the matter fields in the vicinity of the horizon.

\subsection*{A. No firewall}

As it has been shown in {[}28{]} the desired LQFT modification which
arise from the universal coupling between graviton and matter fields
in the near-horizon region $\int\phi^{\mu\nu}T_{\mu\nu}$ need not
stop at $r=2M$ but rather extend beyond the horizon in order to avoid
firewall formation. Giddings argued (see Ref. {[}28{]}) the source
of any such effects need to obey certain constraints

(i) Smooth behavior at $r=2M$

(ii) Rapidly vanishing outside $"the\:zone"$

In what follows we show the conjectured horizon fluctuations {[}13,16{]}
satisfy both conditions.

Let us begin with the latter constraint. The source $(graviton)$
depends on the internal degrees of freedom which can be taken as invariant
on a particular time slice. Thus it is trivial to show the radial-
and time-dependence of the graviton, and how it vanishes rapidly at
large $r$ given the definitions we have provided regarding $weak$
and $strong$ fluctuations. Consider the following gedanken experiment.
Imagine an isolated black hole and an observer coming from past null
infinity who is carrying a measuring apparatus and constantly performing
measurements. Suppose now she gets close to the black hole, and then
accelerates to future null infinity. We wish to know what the radial
dependence of the measurements is. Since the graviton is treated in
terms of the Hawking formalism, we assume the expectation value of
the measurements to rapidly decrease at large $r$. Thus $\left\langle N_{i}\right\rangle =r^{-1}$
with the transition being continuous. That is to say that outside
\textquotedbl{}$the\:zone$\textquotedbl{} $(r\gg2M)$ the graviton
fluctuations are $\ll\lambda$.

The former requirement is satisfied as follows. The non-local effects
arising from the field theory treatment of the graviton are expected
to cause no drama for an observer crossing the horizon since (i) they
have a lifespan of order the fluctuation $\varphi,$ and (ii) are
considered $onto$ the horizon, and are thus spacelike separated from
an observer in Rindler space. Consider the following gedanken experiment.
Suppose we examine the given graviton fluctuations by focusing on
a neighborhood of the Rindler-like horizon, Figure 2. 
\begin{figure}
\includegraphics[scale=0.68]{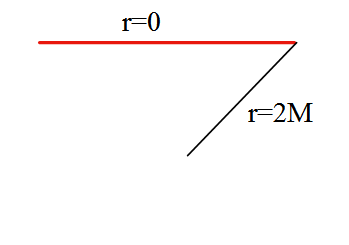}\caption{Diagram of the near-horizon region.}
\end{figure}
 In particular, let us think of the field fluctuations $\varphi$
in terms of harmonic oscillators (HOs). More precisely, imagine we
place harmonic oscillators onto $\mathit{\mathcal{H}}^{+}$ with relatively
small spacing $\epsilon$ in-between

\begin{equation}
\sum_{n=1}^{N}\int_{R}d\varphi(n_{i}\varphi_{i})
\end{equation}
where $R$ denotes the horizon region, and $\varphi$ is the frequency.
The net number of harmonic oscillators is given by $N$, where for
a fixed cut-off $N\sim A_{H}/\epsilon$.

Note the individual HOs need not have the same radial frequency $\varphi$.
The cut-off $\epsilon$ should be taken within an arbitrary distance
spectrum; hence it must be normalized. Again, we are interested in
harmonic oscillators with oscillation frequencies above a certain
threshold $\lambda$. The influence (\textquotedbl{}domain of dependence\textquotedbl{})
of those higher frequency HOs is highly localized, and thus proportional
to their sizes which are considered to be $\gg l_{p}$. As a result,
an individual harmonic oscillator with arbitrary high frequency $cannot$
affect significantly the horizon geometry.

Note the conjectured highly localized and brief violations of locality
occur only in the presence of horizon since it foliates the given
spacetime region into distinct causal patches with continuous transition
between the corresponding CFTs, Eq. (7).

\subsection*{B. Extra quanta}

Restoring the unitary evolution of a black hole in generic models
requires extra quanta to be emitted. Moreover, we wish the Hawking
emission deviations to be consistent with the postulates of observer
complementarity, hence to be no divergence of the stress tensor at
the horizon. Again, we focus on the coupling between graviton and
matter fields in the near-horizon region $\int\phi^{\mu\nu}T_{\mu\nu}$,
and argue that by making certain conservative assumptions, an extra
quantum per time $R$ is achievable in the current framework without
the need of introducing additional degrees of freedom at the horizon
(firewall). Namely, by assuming (i) a black hole begins to decay much
before Page time $(\ll t_{Page})$, (see Ref. {[}16{]}) and (ii) the
extra quantum which carries out the quantum information is of longer
wavelength $(\sim R)$, smooth horizon can be easily achieved even
in the case of an extra particle per $R$. In this case the additional
modes will only appear as a small correction in the overall perturbation.
Note the needed quanta are very few; an extra particle per time $R$
is sufficient to restore the unitary evolution. In {[}16{]} we argued,
by postulating assumptions regarding the internal dynamics of a static
black hole, that quantum information emission in the form of Hawking
modes can begin $\mathcal{O}(R\,logR)$ after the collapse; where
$R\,logR\:\ll t_{Page}$. The additional modes have low Hawking temperature
of $\sim1/R$. Provided information begins leaking $\ll t_{Page}$
{[}16,17{]}, then the entanglement entropy of the black hole will
be negligible compared to the Bekenstein-Hawking bound. Generically,
the extra modes wavelengths are comparable to the horizon radius $\approx R$,
and are thus considered \textquotedbl{}soft.\textquotedbl{} However,
even if the energy density of the additional particles exceeds the
Hawking temperature by a certain factor, there will be still no drama
for an infalling observer. Moreover, the current framework agrees
with the Page spectrum concerning the adiabatic information release
from a black hole which has evaporated less than half of its coarse-grained
entropy. Therefore for a perturbed \textquotedbl{}young\textquotedbl{}
black hole after $R\,log\,R$ the emission rate is 

\begin{equation}
\frac{dI}{dt}\sim exp[-4\pi/y^{2}]
\end{equation}
where $y^{2}=m_{p}/E$, and $E$ is the energy of the emitted radiation. 

\section*{3. Horizon oscillations from weak fluctuations}

In the current Section we focus on the weak fluctuations of the graviton
very near the global horizon, and specifically how they can lead to
the conjectured Planckian-amplitude horizon oscillations {[}13{]}.
In particular, we are interested in the back-reaction of the global
horizon due to fluctuations below the threshold $\lambda$. Considering
the coupling between gravity and matter fields in a black hole spacetime
leads to non-trivial dynamics of $\mathcal{H}^{+}$ due to the \textquotedbl{}amplified\textquotedbl{}
energy density of the fluctuations by the internal degrees of freedom
of the hole. The effect of black hole's mass on the surrounding fields
is best illustrated by the Hawking effect.

Let us begin by defining what we mean by weak fluctuations $(\varphi<\lambda)$.
Similar to the definition of strong fluctuations $(\varphi>\lambda)$
provided in Sec. 2, we define a weak fluctuation of the graviton to
be one that, when considered at asymptotic spatial infinity does not
lead to a particle production. Hence the expectation value of $N_{i}$
vanishes. Suppose Bob carries a very sensitive apparatus which can
detect excitations with arbitrary low energy density, and stays far
away from the black hole. In general, we suspect Bob should measure
negligibly small number for $N_{i}$ compared to $\varphi_{weak}$,
namely $\lambda\gg\left\langle N_{i}\right\rangle $. Since the weak
fluctuations cannot affect significantly the background metric, and
thus the imposed by it locality, they may be thought of as a source
of smaller geometrical disturbances, hence horizon oscillations (fluctuations).
We initially derived the horizon oscillations {[}13{]} from perturbation
theory, and argued they followed the thermodynamic evolution of the
hole. The frequency of the horizon oscillations is given as

\begin{equation}
\omega=\left(\frac{-T_{\mu\nu}}{M_{BH}}\right)^{3/2}
\end{equation}
where $T_{\mu\nu}$ is the stress tensor, or the radiated Hawking
particles.

The oscillations occur naturally in the process of black hole formation/evaporation,
and are expected in any physically-meaningful theory of quantum gravity.
During the evaporation of a black hole we assume 

\begin{equation}
\frac{d^{2}A(t)}{dt^{2}}=-T_{\mu\nu}
\end{equation}
Note that $T_{H}=1/M_{BH}$.

There are certain constraints that the dynamics which take care of
information escape to asymptotic infinity need to respect. If we wish
to keep LQFT in Minkowski space and simultaneously have spacelike
transfer of quantum information a firewall will form. For that reason
Giddings has argued {[}27{]} that if we wish to avoid formation of
a firewall, $T_{\mu\nu}\rightarrow\infty$ as $r\rightarrow2M$, the
modified LQFT must extend beyond the horizon. Thus, the conjectured
horizon fluctuations may serve as the microscopic origin of the desired
effects which normalize the experience of an infalling observer. We
can illustrate the effects of the weak graviton fluctuations on the
background geometry in terms of the strings introduced in Sec. 2.
Consider now a string just outside the global horizon with harmonic
oscillators placed on it, Eq. (12). Since $\varphi_{i}$ are taken
to be below $\lambda$ we expect to see the effects manifest in metric
fluctuations of order the Planck length of the type proposed in {[}13{]}
and in the current paper.

Further, several authors {[}19-23{]} have suggested horizon oscillations
(fluctuations) are generic phenomena in quantum gravity. Let us briefly
comment on the existing literature. First, in Ref. {[}19{]} Bekenstein
and Mukhanov proposed horizon fluctuations may be achieved by describing
the black hole as a quantum system of discrete energy levels. In the
process of decay, the black hole \textquotedbl{}jumps\textquotedbl{}
from one energy level to the next, and thus smears the precise location
of the horizon. Although the approach may seem different than the
classical one we have proposed {[}13{]}, or the currently considered,
the results appear to be identical. Namely, given the particular entropy
spacing between the different energy levels, and the transition between
them, they have derived a frequency equation almost identical to Eq.
(14). As a consequence, beyond a certain threshold for $\omega$,
no radiation will be further emitted {[}24{]}. Furthermore, for a
thermal emission the mass dependence on $T_{\mu\nu}$ appears identical
to Eq. (15). Mathur argued in Ref. {[}29{]} that oscillations of the
horizon surface may be derived from \textquotedbl{}hard impacts\textquotedbl{}
in the context of fuzzball complementarity. More precisely, high energy
infalling quanta $E\gg T$, where $T$ is the local Hawking temperature,
impact the fuzzball surface of the hole and cause oscillations which
wear out and produce low-temperature Hawking particles $E\sim T$
which are supposed to carry out the quantum information. The framework
should be contrasted to the particular paper since the horizon oscillations
in our work result from the generic coupling between the $graviton$
and the fields of nature in Minkowski space. Moreover, the model we
put forward achieves unitary evaporation without diverging the stress
tensor at the horizon. It was shown in {[}30{]} that subtle modifications
of LQFT can indeed lead to emission of quantum information to asymptotic
infinity without the formation of a firewall. For an explicit derivation
of how gravitational collapse of massive shell in a semiclassical
background geometry leads to pure state density matrix see Ref. {[}31{]}.

\section*{4. Conclusions}

By embracing the notion of locality as an effective theory, and by
treating gravity in Minkowski space as a field theory, we presented
a scenario for adiabatic information release from a static black hole
which does not cause drama for an infalling observer and begins much
before Page time. Namely, by introducing extra particle radiation
beyond the Hawking emission we manage to restore the unitary evolution
without forming a firewall. Further, the current framework does not
lead to divergence of the stress-energy tensor at the global horizon
due to the early emission initiation and the low energy density of
the emitted particles. The model presents perturbative quantum effects
which emerge from considering strong fluctuations $(>\lambda)$ of
the gravitational field at the global horizon, namely highly-localized
and brief disturbances of locality, as imposed by the background metric.
On the other hand, by focusing on the graviton fluctuations below
the threshold $\lambda$ very near and at the horizon, we provided
a microscopic origin of the conjectured Planckian-amplitude horizon
oscillations (fluctuations). Therefore, entanglement appears to be
the origin of the horizon fluctuations. Entanglement continuous to
show its importance, not only in the context of emergence of space
and time, but also in quantum gravity phenomena in black holes.
\begin{acknowledgments}
The authors declare that there is no conflict of interest regarding
the publication of this paper.
\end{acknowledgments}

\end{document}